\documentclass[prd,draft,eqsecnum,showpacs]{revtex4}
\usepackage{latexsym}
\begin{document}
\title{Relativistic second-order perturbations of \\ 
nonzero-$\Lambda$ flat cosmological models and CMB anisotropies
}
\author{Kenji Tomita }
\email{tomita@yukawa.kyoto-u.ac.jp}
\affiliation{Yukawa Institute for Theoretical Physics, 
Kyoto University, Kyoto 606-8502, Japan}
\date{\today}

\begin{abstract}
First the second-order perturbations of nonzero-$\Lambda$ cosmological
models are derived explicitly with an arbitrary potential function of 
spatial coordinates, using the nonlinear version of Lifshitz's method
in the synchronous gauge. Their expression is the generalization (to
the nonzero-$\Lambda$ case) of second-order perturbations in the 
Einstein-de Sitter model which were derived previously by the present 
author. Next the second-order temperature anisotropies of Cosmic
Microwave Background radiation are derived using the gauge-invariant
formula which was given by Mollerach and Matarrese. Moreover the
corresponding  
perturbations in the Poisson gauge are derived using the second-order 
gauge transformations formulated by Bruni et al. In the second-order
it is found in spite of gauges that tensor (gravitational-wave)
perturbations and vector (shear) perturbations without
vorticity are induced from the
first-order scalar perturbations. These results will
be useful to analyze the nonlinear effect of local inhomogeneities 
on Cosmic Microwave Background anisotropies.

\end{abstract}

\maketitle


\section{Introduction}
\label{sec:level1}

The anisotropies of the Cosmic Microwave Background radiation 
(CMB) gives us an important information of the structure of our 
universe. So far the relation of CMB anisotropies to inhomogeneities
along light paths has been analyzed using the linear theory of
gravitational instability. Recently the measurements of the 
anisotropies has however become more and more precise, so that
we can catch some information related to nonlinear effect of 
inhomogeneities such as in the form of non-gaussianity \cite{map}.

The general-relativistic second-order nonlinear theory of 
gravitational instability was studied previously by the present author
\cite{tom67,tom71}, in connection with structure 
formation, by extending Lifshitz's linear theory\cite{lifsh} to the
second-order smallness. 
They were restricted to the vanishing $\Lambda$ model (Einstein-de Sitter
model) and the treatment in the synchronous gauge.
Recently the nonlinear gauge transformations and the condition of
second-order gauge-invariantness have been studied by Bruni et
al.\cite{gauge} and the second-order tranformation from the
synchronous gauge to the Poisson gauge has been performed by  
Matarrese et al.\cite{eds}, and the second-order temperature
anisotropy of CMB radiation has been derived by Mollerach and
Matarrese \cite{cmb}. The second-order theory has, moreover, been
extended to useful cases \cite{nong1, nong2, pol, rev}  
of nonzero-$\Lambda$ and non-vanishing pressure, to analyze  
non-gaussianity in the CMB anisotropies.

In this paper we will derive the second-order perturbations
corresponding to the first-order scalar perturbations, in the 
nonzero-$\Lambda$ and pressureless case using the synchronous gauge,
in which 
the perturbations are expressed explicitly with an arbitrary potential 
function ($F$) of spatial coordinates, similarly to our previous work
\cite{tom67}. 

Next we represent the second-order CMB anisotropies
($\mathop{\delta}_2 T/T$) 
using the derived metric perturbations in the synchronous gauge. The
CMB anisotropies, which was derived by Mollerach and Matarrese\cite{cmb} in
arbitrary gauges, are gauge-invariant, so that our expressions will be
useful in the same way as those in the other gauges.

Lastly we derive the second-order perturbations in the Poisson gauge
from those in the synchronous gauge,
using the nonlinear gauge transformation exploited by Bruni et
al.\cite{gauge} and Matarrese et al.\cite{eds} The latter gauge may be
familiar and comprehensive, because 
the line-element in the Poisson gauge is similar to that in the 
Newtonian and Post-Newtonian approximation \cite{new}.

It is found that, in the second-order, tensor (gravitational-wave)
perturbations and vector (shear) perturbations without
vorticity are induced from the first-order scalar perturbations.  

In Appendix A the derivation of basic equations in the extended
version of Lifshitz theory is reviewed.

\section{Second-order perturbations in the synchronous gauge} 
\label{sec:level2}

First our background universe is assumed to be described by isotropic 
and homogeneous pressureless cosmological models which are spatially 
flat, and their spacetimes are expressed by the line-element
\begin{equation}
  \label{eq:m1}
 ds^2 = g_{\mu\nu} dx^\mu dx^\nu = a^2 (\eta) 
[- d \eta^2 + \delta_{ij} dx^i dx^j],
\end{equation}
where the Greek and Latin letters denote $0,1,2,3$ and $1,2,3$,
respectively, contrary to the notation in the previous paper \cite{tom67}, and
$\delta_{ij} (= \delta^i_j = \delta^{ij})$ are the Kronecker
delta. The conformal time $\eta (=x^0)$ is related to the cosmic time
$t$ by $dt = a(\eta) d\eta$.

In the comoving coordinates, the velocity vector and the
energy-momentum tensor of the pressureless matter are expressed as
\begin{equation}
  \label{eq:m2}
u^0 = 1/a, \ \ u^i = 0
\end{equation}
and 
\begin{equation}
  \label{eq:m3}
T^0_0 = -\rho, \ T^0_i = 0, \ T^j_i = 0,
\end{equation}
where $\rho$ is the matter density. From the Einstein equations, the
equations for $\rho$ and the scale factor $a$ are obtained :
\begin{equation}
  \label{eq:m4}
\rho a^2 = 3(a'/a)^2 - \Lambda a^2,
\end{equation}
and
\begin{equation}
  \label{eq:m5}
\rho a^3 = \rho_0,
\end{equation}
where a prime denotes $\partial/\partial \eta$,  $\Lambda$ is the
cosmological constant, and $\rho_0$ is an integration constant.

Next we consider first-order perturbations of the scalar type. The
perturbations of 
metric, matter density and velocity are represented by
$\mathop{\delta}_1 g_{\mu\nu} (\equiv h_{\mu\nu}), \mathop{\delta}_1
\rho,\ {\rm and} \ \mathop{\delta}_1 u^\mu $. 

When we adopt the synchronous coordinates, the metric perturbations
satisfy the condition
\begin{equation}
  \label{eq:m7}
h_{00} = 0 \ {\rm and} \ h_{0i} = 0.
\end{equation}
The scalar-type solutions of the perturbed Einstein equations are classified into
the growing case and the decaying case, and the remaining components
are expressed in both cases as follows:

\bigskip
\noindent (1) the growing case 
\begin{eqnarray}
  \label{eq:m8}
h^j_i &=& \delta^j_i F + P(\eta) F^{|j}_{|i}, \cr
\mathop{\delta}_1 u^0 &=& 0, \ \ \mathop{\delta}_1 u^i = 0, \cr
\mathop{\delta}_1 \rho/\rho &=& {1 \over \rho a^2} \Bigl({a'\over a}P'
-1\Bigr) \Delta F, 
\end{eqnarray}
where $F$ is an arbitrary function of spatial coordinates $x^1, x^2$
and $x^3$, \ $\Delta \equiv \nabla^2$, \ $h^j_i = g^{jl} h_{li}$ and
$P(\eta)$ satisfies  
\begin{equation}
  \label{eq:m9}
P'' + {2a' \over a}P' - 1 =0.
\end{equation}
Its solution is expressed as
\begin{equation}
  \label{eq:m10}
P = \int^\eta_0 d\eta' a^{-2} (\eta') \int^{\eta'}_0 d\eta'' a^2(\eta'').
\end{equation}
The three-dimensional covariant derivatives $|i$ are defined in the
space with metric $dl^2 = \delta_{ij} dx^i dx^j$ and their suffices
are raised and lowered using $\delta_{ij}$, so that their derivatives
are equal to partial derivatives, i.e. $F^{|j}_{|i} = F_{,ij}$, where  
$F_{,i} \equiv  \partial F/\partial x^i$.

\bigskip
\noindent (2) the decaying case
\begin{eqnarray}
  \label{eq:m11}
h^j_i &=&  P(\eta) F_{,ij}, \cr
\mathop{\delta}_1 u^0 &=& 0, \ \ \mathop{\delta}_1 u^i = 0, \cr
\mathop{\delta}_1 \rho/\rho &=& {a'/a \over \rho a^2} P' \Delta F, 
\end{eqnarray}
where $P(\eta)$ satisfies
\begin{equation}
  \label{eq:m12}
P'' + {2a' \over a} P' = 0
\end{equation}
and the solution is
\begin{equation}
  \label{eq:m13}
P = \int^\eta_0 d\eta' a^{-2} (\eta').
\end{equation}
These first-order density perturbations are consistent with the
gauge-invariant variable $\epsilon_m$ (defined by Bardeen\cite{bar}),
which is described by the 
equation 
\begin{equation}
  \label{eq:m14}
{\epsilon_m}'' + {a' \over a} {\epsilon_m}' - {1 \over 2} \rho a^2
\epsilon_m = 0
\end{equation}
in the pressureless case without anisotropic stresses and entropy
perturbations. In fact $\mathop{\delta}_1 \rho/\rho$ in
Eqs.(\ref{eq:m8}) and (\ref{eq:m11}) satisfy Eq. (\ref{eq:m14}) for $P$
in Eqs.(\ref{eq:m10}) and (\ref{eq:m13}).
In both cases, the potential function $F$ is determined by the last
equation of Eq.(\ref{eq:m8}) or Eq.(\ref{eq:m11}) which is regarded as
the cosmological Poisson equation, if the spatial 
distributions of $\mathop{\delta}_1 \rho/\rho$ are given at a
specified epoch. 

Now let us derive the second-order perturbations 
$\mathop{\delta}_2 g_{\mu\nu} (\equiv \ell_{\mu\nu}), \mathop{\delta}_2
\rho,\ {\rm and} \ \mathop{\delta}_2 u^\mu $, where the total
perturbations are 
\begin{eqnarray}
  \label{eq:m15}
\delta g_{\mu\nu} &=& h_{\mu\nu} + \ell_{\mu\nu}, \cr
\delta u^\mu &=& \mathop{\delta}_1 u^\mu +\mathop{\delta}_2 u^\mu, \cr 
\delta \rho/\rho &=& \mathop{\delta}_1 \rho/\rho + \mathop{\delta}_2 \rho/\rho.
\end{eqnarray}
Here from the synchronous gauge condition, we have
\begin{equation}
  \label{eq:m16}
\ell_{00} = 0 \ \ {\rm and} \ \ \ell_{0i} = 0.
\end{equation}
Then, by solving the equations (\ref{eq:a15}) and (\ref{eq:a16}) for
$\ell_{ij}$ which are derived from the 
perturbed Einstein equations in Appendix A, we obtain the following
expressions of $\mathop{\delta}_2 g_{\mu\nu} (\equiv \ell_{\mu\nu})$
and using Eqs. (\ref{eq:a16}) and (\ref{eq:a17}) we obtain $\mathop{\delta}_2 
\rho\ {\rm and} \ \mathop{\delta}_2 u^\mu $.

\bigskip
\noindent (1) the growing case

The metric perturbations reduce to
\begin{equation}
  \label{eq:m17}
\ell^j_i = P(\eta) L^j_i + P^2 (\eta) M^j_i + Q(\eta) N^{|j}_{|i} + C^j_i,
\end{equation} 
where $N^{|j}_{|i} = \delta^{jl} N_{|li} = N_{,ij}$ and $Q(\eta)$ satisfies
\begin{equation}
  \label{eq:m18}
Q'' + {2a' \over a} Q' = P - {5 \over 2} (P')^2
\end{equation} 
and the solution is expressed as
\begin{equation}
  \label{eq:m19}
Q = \int^\eta_0 d\eta' a^{-2}(\eta') \int^{\eta'}_0 d\eta''
a^2(\eta'') \Bigl[P(\eta'') -{5 \over 2}(P'(\eta''))^2\Bigr]. 
\end{equation} 
In the case $\Lambda = 0$, we have \ $P - (5/2)(P')^2 = 0$ because of $a 
\propto \eta^2$ and $P = \eta^2/10$, so that $Q$ vanishes.  
The functions $L^j_i$ and $M^j_i$ are defined by
\begin{eqnarray}
  \label{eq:m20}
L^j_i &=& {1 \over 2}\Bigl[-3 F_{,i} F_{,j} -2 F \cdot F_{,ij} + {1 \over 2}
\delta_{ij} F_{,l} F_{,l}\Bigr], \cr
M^j_i &=& {1 \over 28}\Big\{19F_{,il} F_{,jl} - 12 F_{,ij} \Delta F -
3\delta_{ij} \Bigl[F_{,kl} F_{,kl} -(\Delta F)^2 \Bigr]\Big\}
\end{eqnarray} 
and $N$ is defined by
\begin{equation}
  \label{eq:m21}
\Delta N = {1 \over 28} \Bigl[(\Delta F)^2 - F_{,kl}F_{,kl}\Bigr].
\end{equation} 
The velocity and density perturbations are found to be
\begin{equation}
  \label{eq:m22}
\mathop{\delta}_2 u^0 =0, \ \  \mathop{\delta}_2 u^i =0 
\end{equation} 
and
\begin{eqnarray}
  \label{eq:m23}
\mathop{\delta}_2 \rho/\rho &=& {1 \over 2\rho a^2}\Bigl\{{1 \over
2}(1 - {a' \over a}P') (3F_{,l}F_{,l} + 8F\Delta F) +{1 \over 2}P
[(\Delta F)^2 + F_{,kl}F_{,kl}] \cr
 &+& {1 \over 4}\Bigl[(P')^2 - {2 \over 7}{a'\over a}Q'\Bigr] [(\Delta F)^2 -
F_{,kl}F_{,kl}] - {1 \over 7} {a'\over a}PP' [4F_{,kl}F_{,kl} +
3(\Delta F)^2]  \Big\}.
\end{eqnarray} 
The last term $C^l_i$ satisfies the wave equation
\begin{equation}
  \label{eq:m24}
\Box C^j_i = {3 \over 14}(P/a)^2 G^j_i + {1 \over 7}\Bigl[P - {5 \over 2}
(P')^2 \Bigr] \tilde{G}^j_i,
\end{equation} 
where the operator $\Box$ is defined by
\begin{equation}
  \label{eq:m25}
\Box \phi \equiv g^{\mu\nu} \phi_{;\mu\nu} = -a^{-2}
\Bigl(\partial^2/\partial\eta^2 + {2a' \over a}\partial/\partial \eta -
\Delta \Bigr) \phi
\end{equation} 
for an arbitrary function $\phi$ by use of the four-dimensional covariant
derivative $;$, and $G^j_i$ and $\tilde{G}^j_i$ are expressed as 
\begin{eqnarray}
  \label{eq:m26}
G^j_i &\equiv& \Delta(F_{,ij} \Delta F - F_{,il} F_{,jl}) + (F_{,ij} F_{,kl}
- F_{,ik} F_{,jl})_{,kl} - {1 \over 2}\delta_{ij}\Delta [(\Delta F)^2 -
F_{,kl}F_{,kl}], \cr
\tilde{G}^j_i &\equiv& F_{,ij} \Delta F - F_{,il} F_{,jl} - {1 \over
4}\delta_{ij}[(\Delta F)^2 - F_{,kl}F_{,kl}] - 7 N_{,ij}. 
\end{eqnarray}
These functions satisfy the traceless and transverse relations
\begin{eqnarray}
  \label{eq:m27}
G^l_l = 0, && G^l_{i,l} = 0, \cr
\tilde{G}^l_l = 0, && \tilde{G}^l_{i,l} = 0,
\end{eqnarray}
so that $C^j_i$ also satifies
\begin{equation}
  \label{eq:m28}
C^l_l = 0, \ \ C^l_{i,l} = 0.
\end{equation} 
This means that $C^j_i$ represents the second-order gravitational
radiation emitted by first-order density perturbations.
The solution of the above inhomogeneous wave equation
(Eq.(\ref{eq:m24})) can be represented in an explicit form using 
the retarded Green function for the operator $\Box$ \cite{tom67,brem,nk}.  
\bigskip

\noindent (2) the decaying case

The metric perturbations are
\begin{equation}
  \label{eq:m29}
\ell^j_i = {1 \over 4} P^2(\eta) \Big\{2F_{,il}F_{,jl} -\Delta F \cdot F_{,ij}
+ {1 \over 4}\delta_{ij} [(\Delta F)^2 - F_{,kl}F_{,kl}] \Big\} + C^j_i,
\end{equation} 
where $P(\eta)$ in this case is given by Eq.(\ref{eq:m13}). The last term $C^j_i$
is described by the wave equation
\begin{equation}
  \label{eq:m30}
\Box C^j_i = -{1 \over 8} (P/a)^2 G^j_i,
\end{equation} 
where $G^j_i$ satisfies Eq.(\ref{eq:m26}).

The velocity and density perturbations are
\begin{equation}
  \label{eq:m31}
\mathop{\delta}_2 u^0 = 0, \ \mathop{\delta}_2 u^i = 0 
\end{equation} 
and
\begin{equation}
  \label{eq:m32}
\mathop{\delta}_2\rho/\rho = {1 \over 8\rho a^2} P' \Big\{P'[(\Delta
F)^2 - F_{,kl}F_{,kl}] - {a'\over a}P[(\Delta F)^2 +3 F_{,kl}F_{,kl}]
\Big\}.
\end{equation}  
Here $\mathop{\delta}_2\rho/\rho$ is the matter density
perturbations observed by comoving observers and $C^j_i$ represents
the gravitational radiation emitted by the first-order density
perturbations. 

By the way we consider the rotational velocity $\omega^\alpha$ and the
corresponding scalar quantity $\omega$ defined by
\begin{equation}
  \label{eq:m32a}
\omega^\alpha \equiv {1 \over 2} \eta^{\alpha\beta\mu\nu} u_\beta
u_{\mu,\nu} \ \ {\rm and } \ \ \omega \equiv (\omega^\alpha \omega_\alpha)^{1/2}.
\end{equation}  
In the above perturbations, it is clear that $\omega^\alpha$ vanishes,
because $\mathop{\delta}_1 u^\mu = \mathop{\delta}_2 u^\mu = 0$, and
so $\omega$ also vanishes.  

Next, for a later use we here express our solutions (in the growing
case) using the notation which was employed by Matarrese, Mollerach and
Bruni \cite{eds, cmb} for the gauge-invariant treatment of second-order
perturbations. In their notation our perturbations are expressed in
the following form:
\begin{eqnarray}
  \label{eq:m33}
g_{00} &=& -a^2(\eta)[1 + 2\psi^{(1)} + \psi^{(2)}], \cr
g_{0i} &=& a^2(\eta)\Bigl[z^{(1)}_{i} + {1 \over 2} z^{(2)}_{i}\Bigr], \cr
g_{ij} &=& a^2(\eta)\Big\{[1 - 2\phi^{(1)} - \phi^{(2)}] \delta_{ij} +
\chi^{(1)}_{ij} + {1 \over 2}\chi^{(2)}_{ij} \Big\}, 
\end{eqnarray}
and 
\begin{equation}
  \label{eq:m34}
\chi^{(r)}_{ij} = \chi^{(r)\|}_{ij} + \chi^{(r)\top}_{ij}, \ \ (r = 1, 2),
\end{equation} 
where $\|$ and $\top$ denote the
scalar and tensor perturbations, respectively, and
\begin{equation}
  \label{eq:m35}
D_{ij} \equiv \partial_i \partial_j - {1 \over 3} \delta_{ij} \Delta.
\end{equation} 
The velocity and density perturbations are
\begin{equation}
  \label{eq:m36}
u^\mu = {1 \over a} \Bigl[\delta^\mu_0 + v^\mu_{(1)} + {1 \over 2}
v^\mu_{(2)}\Bigr] 
\end{equation} 
and
\begin{equation}
  \label{eq:m37}
\rho = \rho_{(0)} + \delta^{(1)} \rho + {1 \over 2} \delta^{(2)} \rho.
\end{equation} 
For our perturbations in the synchronous gauge, we have
\begin{equation}
  \label{eq:m38}
\psi^{(1)} = \psi^{(2)} = z^{(1)}_i = z^{(2)}_i = 0
\end{equation} 
and the other components are expressed by use of our notation as
\begin{eqnarray}
  \label{eq:m39}
\phi^{(1)} &=& -h/6  = -{1 \over 2} (F + {1 \over 3} P \Delta F), \cr
\phi^{(2)} &=& - \ell/3 = - {1 \over 3} (P L^i_i + P^2 M^i_i + Q \Delta N),\cr
\chi^{(1)\|}_{ij} &=& D_{ij} \chi^{(1)\|} = h^j_i - {1 \over 3}
\delta^j_i h = P \Bigl(F_{,ij} - 
{1 \over 3} \delta _{ij} \Delta F\Bigr) \ \ {\rm or} \ \ \chi^{(1)\|}
= P F, \cr 
\chi^{(1)\top}_{ij} &=& 0, \cr
{1 \over 2} \chi^{(2)\|}_{ij} &=& \ell^j_i - {1 \over 3} \delta^j_i
\ell = P L^j_i +P^2   
M^j_i + Q N_{,ij} -{1 \over 3} \delta _{ij} (P L^k_k +P^2 M^k_k +
Q\Delta N), \cr
{1 \over 2} \chi_{ij}^{(2)\top} &=& C^j_i, \cr
v^\mu_{(1)} &=& v^\mu_{(2)} = 0, \cr
\delta^1 \rho &=& \mathop{\delta}_1 \rho, \ \ {1 \over 2} \delta^{(2)} \rho =
\mathop{\delta}_2 \rho, 
\end{eqnarray}
where $h = h^k_k$ and $\ell = \ell^k_k$.   It is interesting that
$\chi^{(2)\|}_{ij}$ includes not only scalar perturbations, but also
vector (shear) perturbations (without vorticity \cite{bar}), 
because it does not reduce to the form of $D_{ij} \chi$ in general.

\section{CMB anisotropies due to first-order and second-order 
perturbations}
\label{sec:level3}

In the unperturbed model universe, the observed temperature ($T_0$) of
the CMB radiation is related to the emitted temperature ($T_e$) at the
decoupling epoch ($z_e$) by $T_o = T_e/(1 + z_e)$, and represented
also as 
\begin{equation}
  \label{eq:b1}
T_o = (\omega_o/\omega_e) T_e
\end{equation}
using the observed and emitted frequencies $\omega_o$ and $\omega_e (=
(1 + z_e) \omega_o$).

In the perturbed universe, these temperatures depend on the motions of
matter and observers and on light paths passing through the
inhomogeneous matter, and are expressed as
\begin{equation}
  \label{eq:b2}
T_o ({\bf x}_o, {\bf e}) = (\omega_o/\omega_e) T_e ({\bf p}, {\bf d})
\end{equation}
with $\omega = -g_{\mu\nu} u^\mu k^\nu$, where $u^\mu$ is the
observer's and emitter's velocities, \ $k^\nu (= dx^\nu/d\lambda)$ is
the wave vector of photons with affine parameter $\lambda$, $({\bf
x}_o, {\bf e})$ and $({\bf p}, {\bf d})$ are (the position vectors and
directional unit vectors) of the observer and emitter, respectively.

The wave vector $k^\mu$ satisfies the perturbed null geodesic equation
and its solutions to the second-order are expressed as
\begin{eqnarray}
  \label{eq:b3}
k^\mu &=& k_{(0)}^\mu + k_{(1)}^\mu + k_{(2)}^\mu, \cr
x^\mu &=& x_{(0)}^\mu + x_{(1)}^\mu + x_{(2)}^\mu,
\end{eqnarray}
where $x^\mu$ represents the light path and $(r)$ denotes the
$r$-order smallness.

The temperature at the decoupling epoch is expressed as
\begin{equation}
  \label{eq:b4}
T_e ({\bf p}, {\bf d}) = T_e^{(0)} [1 + \tau ({\bf p}, {\bf d})]
\end{equation}
and the frequencies to the second-order are
\begin{equation}
  \label{eq:b5}
\omega = \omega^{(0)} [1 + \tilde{\omega}^{(1)} + \tilde{\omega}^{(2)}]. 
\end{equation}
Then we have
\begin{equation}
  \label{eq:b6}
T_o ({\bf x}_o, {\bf e}) = T_o^{(0)} [1 + \mathop{\delta}_1 T/T +
\mathop{\delta}_2 T/T], 
\end{equation}
where 
\begin{eqnarray}
  \label{eq:b7}
\mathop{\delta}_1 T/T &=& \tilde{\omega}^{(1)}_o -
\tilde{\omega}^{(1)}_e + \tau, \cr
\mathop{\delta}_2 T/T &=& \tilde{\omega}^{(2)}_o -
\tilde{\omega}^{(2)}_e + (\tilde{\omega}^{(1)}_e)^2 -
\tilde{\omega}^{(1)}_o \tilde{\omega}^{(1)}_e +
(\tilde{\omega}^{(1)}_o - \tilde{\omega}^{(1)}_e)\tau +
p^{(1)i}\partial \tau/\partial x^i + d^{(1)i} \partial \tau/\partial d^i.
\end{eqnarray}
The procedure for solving null geodesic equations in perturbed
universe models was shown by Pyne and Carroll, in which 
the background null rays are given by $x^{(0)\mu} = (\lambda,
(\lambda_o - \lambda)e^i)$ and $k^{(0)\mu} = (1, -e^i)$, and the
baoundary conditions at the origin are $x^{(1)\mu}(\lambda_o) =
x^{(2)\mu}(\lambda_o) =0$ and $k^{(1)i}(\lambda_o)
=k^{(2)i}(\lambda_o) =0$.  The expressions for 
$\mathop{\delta}_1 T/T$ and $\mathop{\delta}_2 T/T$ were derived by
Mollerach and Matarrese \cite{cmb} in general gauges using Pyne and Carroll's
procedure \cite{pyn}. Their expressions are found to be gauge-invariant, and so the
values can be calculated in a specified gauge without loss of any
generality. Here we describe them using 
our solutions in the synchronous gauge, under the condition that the
observers and 
emitters are comoving, i.e. $v^{(r)}_o = v^{(r)}_e = 0$.

The following temperature perturbations are derived in the synchronous
gauge from equations  
(2.20) - (2.28) in Mollerach and Matarrese's paper \cite{cmb}.
First we have the first-order perturbations: 
\begin{equation}
  \label{eq:b8}
\mathop{\delta}_1 T/T =  \tau - I_1 (\lambda_e), 
\end{equation}
where
\begin{equation}
  \label{eq:b9}
I_1(\lambda) = \int^\lambda_{\lambda_o} d\bar{\lambda}
{A^{(1)}}'(\bar{\lambda}),  
\end{equation}
\begin{equation}
  \label{eq:b10}
A^{(1)} = \phi^{(1)} - {1 \over 2} \chi^{(1)}_{ij} e^i e^j, 
\end{equation}
and $e^i$ denotes a component of the directional unit vector ${\bf e}$.
The first-order wave vectors are
\begin{equation}
  \label{eq:b11}
k^{(1)0}(\lambda) =  - \phi^{(1)}_o + {1 \over 2}\chi^{(1)ij}_o e_i e_j + I_1
(\lambda)  
\end{equation}
and 
\begin{equation}
  \label{eq:b12}
k^{(1)i}(\lambda) = 2\phi^{(1)}_o e^i - \chi^{(1)ij}_o e_j
- 2\phi^{(1)} e^i + \chi^{(1)ij} e_j - I_1^i (\lambda),
\end{equation}
where
\begin{equation}
  \label{eq:b13}
I_1^i (\lambda) = \int^\lambda_{\lambda_o} d\bar{\lambda} A^{(1)|i}
(\bar{\lambda}) 
\end{equation}
and $|i = \delta^{il} \partial_l$ and $e_j = \delta_{ij} e^i$. 

The first-order light paths are
\begin{eqnarray}
  \label{eq:b14}
x^{(1)0} (\lambda) &=& (\lambda -\lambda_o) \Bigl[- \phi^{(1)}_o + {1
\over 2}\chi^{(1)ij}_o e_i e_j\Bigr] \cr
&+& \int^\lambda_{\lambda_o} d\bar{\lambda} (\lambda -\bar{\lambda})
{A^{(1)}}'(\bar{\lambda}),    
\end{eqnarray}
\begin{eqnarray}
  \label{eq:b15}
x^{(1)i} (\lambda) &=& (\lambda -\lambda_o)[2\phi^{(1)}_o e^i  -
\chi^{(1)ij}_o  e_j] \cr 
&-& \int^\lambda_{\lambda_o} d\bar{\lambda} [2\phi^{(1)} e^i 
-\chi^{(1)ij} e_j +(\lambda -\bar{\lambda}) A^{(1)|i}(\bar{\lambda})].
\end{eqnarray}
The second-order wave vectors satisfy the following relation
\begin{equation}
  \label{eq:b16}
k^{(2)0}_e - k^{(2)0}_o =  I_2 (\lambda_e),
\end{equation}
where
\begin{equation}
  \label{eq:b17}
I_2(\lambda) = \int^\lambda_{\lambda_o} d\bar{\lambda} \Bigl[{1 \over
2} {A^{(2)}}' + {\chi^{(1)}_{ij}}' e^j(k^{(1)i} +
e^i k^{(1)0}) +2k^{(1)0} {A^{(1)}}' + 2{\phi^{(1)}}' A^{(1)} +
x^{(1)0}{A^{(1)}}''  + x^{(1)i}{A^{(1)}_{,i}}' \Bigr]
\end{equation}
and
\begin{equation}
  \label{eq:b18}
A^{(2)} \equiv  \phi^{(2)} - {1 \over 2}{\chi^{(2)}_{ij}} e^i e^j.
\end{equation}
The second-order temperature perturbations are
\begin{eqnarray}
  \label{eq:b19}
\mathop{\delta}_2 T/T &=& - I_2 (\lambda_e) + I_1 (\lambda_e) \Bigl[I_1
(\lambda_e) -\tau - \phi^{(1)}_o +{1 \over 2}\chi^{(1)ij}_o e_i
e_j \Bigr] \cr
&+& x^{(1)0}_e {A^{(1)}_e}' + (x^{(1)j}_e +x^{(1)0}_e e^j)\tau_{,j} +
{\partial\tau \over \partial d^i} d^{(1)i},
\end{eqnarray}
where
\begin{equation}
  \label{eq:b20}
d^{(1)i} = e^i - (e^i - k^{(1)i})/|e^i - k^{(1)i}|.
\end{equation}

Moreover, if we substitute our metric perturbations into the above
equations, we obtain
\begin{equation}
  \label{eq:b21}
\mathop{\delta}_1 T/T =  \tau + {1 \over 2}\int^{\lambda_e}_{\lambda_o}
d\lambda P'(\eta) F_{,ij} e^i e^j
\end{equation}
for the first-order perturbation. Since $dP/d\lambda = P'$ and
$dF/d\lambda = - F_{,i} e^i$, Eq.(\ref{eq:b21}) can be expressed as
\begin{equation}
  \label{eq:b22}
\mathop{\delta}_1 T/T =  \Theta_1 +\Theta_2
\end{equation}
with 
\begin{eqnarray}
  \label{eq:b23}
\Theta_1 &\equiv& \tau - {1 \over 2}[(P' F_{,i})_e - (P' F_{,i})_o] e^i, \cr
\Theta_2 &\equiv& {1 \over 2}\int^{\lambda_e}_{\lambda_o}
d\lambda P''(\eta) F_{,i} e^i.
\end{eqnarray}
The latter term $\Theta_2$ represents the first-order Sachs-Wolfe
effect.

Next, we notice that Eqs.(\ref{eq:b11}), (\ref{eq:b12}), (\ref{eq:b14}) and
(\ref{eq:b15}) lead to 
\begin{eqnarray}
  \label{eq:b24}
k^{(1)0} + k^{(1)i}e_i &=& - A^{(1)}, \cr
x^{(1)0} + x^{(1)i}e_i &=& - \int^{\lambda}_{\lambda_o} d\lambda A^{(1)},
\end{eqnarray}
and we have
\begin{equation}
  \label{eq:b24a}
\int^{\lambda_e}_{\lambda_o} d\lambda {A^{(1)}}' (\lambda) I_1 (\lambda)
= {1 \over 2} [I_1 (\lambda_e)]^2.
\end{equation}
Then we obtain from Eq.(\ref{eq:b17})
\begin{eqnarray}
  \label{eq:b25}
I_2 (\lambda_e) &=& {1 \over 2} [I_1 (\lambda_e)]^2 - (\lambda_e -
\lambda_0) A^{(1)}_o  {A^{(1)}_e}' +
{A^{(1)}_e}' \int^{\lambda_e}_{\lambda_o} d\lambda [A^{(1)} +(\lambda_e -
\lambda) {A^{(1)}}']  \cr 
&-&  A^{(1)}_o I_1 (\lambda_e) + \int^{\lambda_e}_{\lambda_o} d\lambda
\Big\{{1 \over 2}{A^{(2)}}' + A^{(1)}{A^{(1)}}' - {A^{(1)}}''
\int^{\lambda}_{\lambda_o} d\bar{\lambda} A^{(1)}(\bar{\lambda}) \Big\},
\end{eqnarray}
and therefore from Eq.(\ref{eq:b19}) 
\begin{eqnarray}
  \label{eq:b26}
\mathop{\delta}_2 T/T &=& I_1 (\lambda_e) \Bigl[{1 \over 2} I_1 (\lambda_e)
- \tau\Bigr] - [{A^{(1)}_e}' + \tau_{,i} e^i] \int^{\lambda_e}_{\lambda_o}
d{\lambda} A^{(1)} \cr
&-& \int^{\lambda_e}_{\lambda_o} d{\lambda} \Big\{{1 \over 2} {A^{(2)}}' 
+ A^{(1)}{A^{(1)}}' - {A^{(1)}}'' 
\int^{\lambda}_{\lambda_o} d\bar{\lambda} A^{(1)}(\bar{\lambda})
\Big\} + {\partial\tau \over \partial d^i} d^{(1)i},
\end{eqnarray}
where $(\eta, x^i) = (\lambda, \lambda_o - \lambda)$ in the 
integrands and
\begin{eqnarray}
  \label{eq:b27}
I_1 (\lambda_e) &=& - {1 \over 2} \int^{\lambda_e}_{\lambda_o}
d{\lambda} P' F_{,ij} e^i e^j, \cr
A^{(1)} &=& - {1 \over 2}  P F_{,ij} e^i e^j , \cr
A^{(2)} &=& - {1 \over 2} [P L^j_i + P^2 M^j_i + Q N_{,ij} +
C^j_i ] e^i e^j. 
\end{eqnarray}
In Eq.(\ref{eq:b26}) the terms with path integrations represent the
second-order nonlinear Integral Sachs-Wolfe effect, which brings the
observational coupling of two linearly independent inhomogeneities
with different wavelengths. As can be seen from Eq.(\ref{eq:b27}), the
induced gravitational radiation contributes to the CMB anisotropies.

\section{Perturbations in the Poisson gauge}
\label{sec:level4}

In this section we derive the perturbations in the Poisson gauge
which is defined by the condition
\begin{equation}
  \label{eq:c1}
z_i^{(r)|i} = 0 \ {\rm and } \ \chi_{ij}^{(r)|j} = 0.
\end{equation}
For this purpose, we use a gauge transformation from the
perturbations in the synchronous gauge to those in Poisson gauge. The
first-order gauge transformation has fully been studied by many
authors (Bardeen \cite{bar}, Kodama and Sasaki \cite{ks}). The second-order
gauge transformation has more recently been derived by Bruni et
al.\cite{gauge} and the transformations of an arbitrary perturbed tensor 
${\cal T} = {\cal T}_0 + \delta {\cal T}^{(1)} + {1 \over 2} \delta 
{\cal T}^{(2)}$ are expressed
in terms of generators $\xi_{(1)}$ and $\xi_{(2)}$ as
\begin{eqnarray}
  \label{eq:c2}
\delta \bar{\cal T}^{(1)} &=& \delta {\cal T}^{(1)} + 
{\cal{L}}_{\xi_{(1)}} {\cal T}_0, \cr
\delta \bar{\cal T}^{(2)} &=& \delta {\cal T}^{(2)} + 
2 {\cal{L}}_{\xi_{(1)}} \delta
{\cal T}^{(1)} + {\cal{L}}_{\xi_{(1)}}^2 {\cal T}_0 + {\cal{L}}_{\xi_{(2)}} 
{\cal T}_0,
\end{eqnarray}
where $\cal{L}$ denotes the Lie derivative and the components of the
generators are expressed as
\begin{equation}
  \label{eq:c3}
\xi^0_{(r)} = \alpha^{(r)}
\end{equation}
and 
\begin{equation}
  \label{eq:c4}
\xi^i_{(r)} = \beta^{(r)|i} + d^{(r)i}
\end{equation}
with ${d^{(r)i}}_{,i} = 0$.

This gauge transformation has been applied by Matarrese et
al. to derive the second-order perturbations of 
the Einstein-de Sitter model in the Poisson gauge from those in the
synchronous gauge. Here in the similar manner we will apply the
transformation to our perturbations derived in \S 2.

\subsection{First-order transformation}
For the transformation of our metric perturbations from the
synchronous gauge (S) to the Poisson gauge (P), we have
\begin{eqnarray}
  \label{eq:c5}
\psi_P^{(1)} &=& \alpha'^{(1)} + {a' \over a} \alpha^{(1)}, \cr
\alpha^{(1)} &=& \beta'^{(1)}, \cr
z^{(1)}_{Pi} &=& {d^{(1)}_i}', \cr
\phi_P^{(1)} &=& \phi_S^{(1)} - {1 \over 3} \Delta \beta^{(1)} - {a'
\over a} \alpha^{(1)}, \cr
\chi_S^{(1)\|} &+& 2 \beta^{(1)} = 0, \cr
d^{(1)}_i &=& 0.
\end{eqnarray}
Using Eq.(\ref{eq:m39}) for $\phi_S^{(1)}$ and $\chi_S^{(1)\|}$, we obtain
from the above equations 
\begin{eqnarray}
  \label{eq:c6}
\alpha^{(1)} &=& - {1 \over 2} P' F, \cr
\beta^{(1)} &=& - {1 \over 2} P F, \cr
\psi_P^{(1)} &=& \phi_P^{(1)} = {1 \over 2} \Bigl(1 - {a'\over a} P'\Bigr) F, \cr
z^{(1)}_{Pi} &=& 0.
\end{eqnarray}

The density and velocity perturbations satisfy the following
relations:
\begin{eqnarray}
  \label{eq:c7}
\delta \rho^{(1)}_P/\rho &=& - \delta \rho^{(1)}_S/\rho  + {\rho' \over \rho} \alpha^{(1)}, \cr
v^{(1)0}_P &=& - {a' \over a} \alpha^{(1)} - \alpha'^{(1)}, \cr
v^{(1)i}_P &=&  - \beta'^{(1)|i} - d'^{(1)i},
\end{eqnarray}
so that 
\begin{eqnarray}
  \label{eq:c8}
\delta \rho^{(1)}_P/\rho &=& {1 \over 2\rho a^2} \Bigl({a' \over a}P'
-1\Bigr) \Delta F + {1 \over 2}{a' \over a} P' F, \cr
v^{(1)0}_P &=& - {1 \over 2}{a' \over a}\Bigl({a' \over a}P' -1\Bigr) F, \cr
v^{(1)i}_P &=&  {1 \over 2} P' F_{,i}.
\end{eqnarray}

\subsection{Second-order transformation}
Similarly we use the transformations expressed as
\begin{eqnarray}
  \label{eq:c9}
\delta g_{\mu\nu P}^{(2)} &=& \delta g_{\mu\nu S}^{(2)} + 
2 {\cal{L}}_{\xi_{(1)}} \delta g_{\mu\nu S}^{(1)} +
{\cal{L}}_{\xi_{(1)}}^2 
g_{\mu\nu}^{(0)} + {\cal{L}}_{\xi_{(2)}} g_{\mu\nu}^{(0)}, \cr
\delta \rho^{(2)}_P &=& \delta \rho^{(2)}_S +
({\cal{L}}_{\xi_{(2)}}+ {\cal{L}}_{\xi_{(1)}}^2) \rho_{(0)} + 2
{\cal{L}}_{\xi_{(1)}} \delta \rho_S^{(1)}, \cr
\delta u^\mu_{(2)P} &=& \delta u^\mu_{(2)S} +
({\cal{L}}_{\xi_{(2)}}+ {\cal{L}}_{\xi_{(1)}}^2) u^\mu_{(0)} + 2
{\cal{L}}_{\xi_{(1)}} \delta u^\mu_{(1)S}. 
\end{eqnarray}
More explicit expressions for these relations are given in Eqs. (5.8) -
(5.14) of Matarrese et al.'s paper \cite{eds}. 
By analyzing these latter equations, we obtain the following generators
\begin{eqnarray}
  \label{eq:c11}
\alpha^{(2)} &=& P' \Bigl({50 \over 3} \Theta_0 + {1 \over 2} F^2\Bigr) + {50 \over 3}
P' P'' \Theta_0 - {100 \over 21}\Bigl(PP' - {1 \over 6} Q'\Bigr) \Psi_0, \cr
\beta^{(2)} &=& P \Bigl({50 \over 3} \Theta_0 + F^2\Bigr) + (P')^2 \Bigl({25 \over 3}
\Theta_0 +{1 \over 8} F^2\Bigr) + P^2 \Bigl(-{50 \over 21} \Psi_0 + {1 \over 8}
F_{,l}F_{,l}\Bigr) + {50 \over 63} Q \Psi_0, \cr 
\Delta d^{(2)}_i &=& \Bigl[- {200 \over 9} \Psi_{0,i} + {1 \over
2}(F_{,l}F_{,l})_{,i} - F_{,i} \Delta F\Bigr]\Bigl[P + {1 \over 2} (P')^2\Bigr],
\end{eqnarray}
where
\begin{eqnarray}
  \label{eq:c12}
\Psi_0 &\equiv& {9 \over 200} [F_{,kl}F_{,kl} - (\Delta F)^2],  \cr
\Delta \Theta_0 &\equiv&  \Psi_0 - {3 \over 100} F_{,l}F_{,l}.
\end{eqnarray}
It is found from Eq.(\ref{eq:c11}) that vector perturbations without
vorticity\cite{bar} appear also in this gauge. 
The resulting metric perturbations are expressed as
\begin{eqnarray}
  \label{eq:c13}
\psi_P^{(2)} &=& {1 \over 4}\Big\{4 - 7{a' \over a}P' +\Bigl[-{a'' \over
a} + 5\Bigl({a' \over a}\Bigr)^2\Bigr](P')^2 \Big\}F^2 + {1 \over 4}P\Bigl(1 - {a'
\over a}P'\Bigr) F_{,l}F_{,l} \cr
&+& {50 \over 3}\Bigl\{2 - {6a' \over a}P' + \Bigl[-{2a'' \over a} + 8\Bigl({a'
\over a}\Bigr)^2 \Bigr](P')^2 \Bigr\} \Theta_0 - {100 \over 21}\Bigl[(P')^2
+ P\Bigl(1 - {a' \over a}P'\Bigr) \cr
&-& {1 \over 6}\Bigl(Q''+ {a' \over a}Q'\Bigr)\Bigr] \Psi_0,
\end{eqnarray}
\begin{eqnarray}
  \label{eq:c14}
\phi_P^{(2)} &=& {1 \over 4}P'\Big\{{a' \over a}+ \Bigl[-{a'' \over
a}+\Bigl({a' \over a}\Bigr)^2 \Bigr]P' \Big\} F^2 + {1 \over 4}P \Bigl(1 - {a'
\over a}P'\Bigr) F_{,l}F_{,l} \cr 
&-& {100 \over 3} {a' \over a}P'\Bigl(1 - {a' \over a}P'\Bigr) \Theta_0 +
\Big\{{100 \over 21} {a' \over a}\Bigl(PP' - {1 \over 6}Q'\Bigr) - {50 \over
9}\Bigl[P + {1 \over 2}(P')^2\Bigr] \Big\} \Psi_0,
\end{eqnarray}
\begin{equation}
  \label{eq:c15}
(\Delta z_i^{(2)})_P = P' (1 + P'') \Bigl[- {200 \over 9} \Psi_{0,i} + {1
\over 2} (F_{,l}F_{,l})_{,i} - F_{,i} \Delta F\Bigr],
\end{equation}
and
\begin{equation}
  \label{eq:c16}
(\chi_{ij}^{(2)\|})_P = 0 \ \ {\rm and} \ \ {1 \over 2}
(\chi_{ij}^{(2)\top})_P = C_{ij},  
\end{equation}
where we used Eq.(\ref{eq:m9}) in the derivation of the above
equations, and $\|$ and $\top$ denote the scalar perturbation and the
transverse and traceless part representing induced gravitational
radiation, respectively. In the limit of $\Lambda = 0$, these
generators and solutions reduce to Eqs.(6.6), (6.7) and (6.8) in
the Matarrese et al.'s paper \cite{eds}, which were shown for the
Einstein-de Sitter model.  

The density and velocity perturbations are
\begin{eqnarray}
  \label{eq:c17}
(\delta^{(2)} \rho/\rho)_P &=& -100 {a' \over a}P' \Bigl(1 -{a' \over
a}P'\Bigr)\Theta_0 
+ {100 \over 7}{a' \over a} \Bigl(PP' - {1\over 6} Q'\Bigr) \Psi_0 +
\Big\{3\Bigl[\Bigl({a' \over a}\Bigr)^2 - {1 \over 4}{a'' \over a}
\Bigr](P')^2 \cr 
&-& {3 \over 4}{a' \over a}P'(P'' +2) \Big\} F^2 
+ {3 \over 2}\Bigl[-2{a' \over a}PP' + (1 - {a' \over a}P')/(\rho
a^2)\Bigr] F_{,l}F_{,l} \cr
&+& {1 \over \rho a^2}\Big\{ -{1 \over 2} P' \Bigl[\Bigl({a'' \over a} -
5({a' \over a})^2\Bigr)P' + 3{a' \over a}\Bigr] + 4\Bigl(1 - {a' \over
a}P'\Bigr)\Big\} F \Delta F \cr
&+& {1 \over \rho a^2} \Big\{{1 \over 14} {a' \over a}Q' [(\Delta F)^2 -
F_{,kl}F_{,kl}] -{1 \over 2} P \Bigl({a' \over a}P' -1\Bigr) F_{,l}
\Delta F_{,l} \cr
&-& {1 \over 7}{a' \over a} PP' [3(\Delta F)^2 + 4
F_{,kl}F_{,kl}] + {1\over 2} P [(\Delta F)^2 + F_{,kl}F_{,kl}] \cr
&+& {1 \over 4} (P')^2 [(\Delta F)^2 - F_{,kl}F_{,kl}] \Big\},  
\end{eqnarray}
\begin{eqnarray}
  \label{eq:c18}
(v^0_{(2)})_P &=& - {100 \over 3}\Bigl[1 - 3{a' \over a}P' +
2\Bigl({a' \over a}\Bigr)^2 
(P')^2\Bigr] \Theta_0 + {100 \over 21} \Bigl[(P')^2 + P\Bigl(1 - {a'
\over a}P'\Bigr) \cr
&+& {1 \over 6}\Bigl(Q'' + {a' \over a}Q'\Bigr)\Bigr] \Psi_0 
+ {1 \over 4} \Bigl[-1 + {a' \over a}P' + \Bigl({a'' \over a} - 2({a' \over
a})^2\Bigr)(P')^2\Bigr] F^2 \cr
&+& {1 \over 4} \Bigl[(P')^2 - P \Bigl(1 -
{a' \over a}P'\Bigr)\Bigr] F_{,l} F_{,l}. 
\end{eqnarray}
\begin{eqnarray}
  \label{eq:c19}
(v^i_{(2)})_P &=& -{d^i_{(2)P}}' - {50 \over 3} P' (1 + P'') \Theta_{0,i} +
{100 \over 21} \Bigl(PP' - {1 \over 6} Q'\Bigr) \Psi_{0,i} \cr
&-& {1 \over 2} PP' F_{,l} F_{,li} - {1 \over 2} P'\Bigl(5 - 3{a' \over
a}P'\Bigr) F F_{,i}.  
\end{eqnarray}
In the limit of $\Lambda = 0$, these relations also are consistent with
Eqs. (6.10), (6.11) and (6.12) in Matarrese et al.'s paper \cite{eds},
except for a few terms which may 
include some misprints with respect to $(\delta^{(2)} \rho)_P$ and
$(v^i_{(2)})_P$.   

Now we consider the second-order rotational velocity in the Poisson
gauge. The rotational velocity vector ($\omega^{\alpha (r)}_P$) and
the corresponding scalar quantity ($\omega^{(r)}_P$)
 in the Poisson gauge are related to those (S) in the
synchronous gauge by Eq.(\ref{eq:c2}) as $\delta \bar{\cal T}^{(r)} =
\omega^{\alpha (r)}_P, \omega^{(r)}_P$ and $\delta {\cal T}^{(r)} =
\omega^{\alpha (r)}_S, \omega^{(r)}_S$, respectively, for $r = 1, 2$.
Here we have ${\cal T}^0 = 0$ and $\delta {\cal T}^{(r)} = 0$ for $r =
1, 2$, as was 
shown in \S 2. It is found therefore from these relations that
$\omega^{\alpha (r)}_P$ and $\omega^{(r)}_P$ vanish. That is, no
rotational perturbations are induced also in the Poisson
gauge. However vector (shear) perturbations without vorticity appear in the
form of $\Delta d^{(2)}_i$.  
 
\section{Concluding remarks}

We have studied the second-order perturbations in spatially flat,
pressureless cosmological models with nonzero cosmological constant,
and could describe them explicitly using an arbitrary potential
function $F$ of spatial
coordinates. It seems however to be difficult to 
derive similar results in the other cases with finite pressures.

It is found in the second-order that tensor (gravitational-wave)
perturbations and vector (shear) perturbations without vorticity are 
induced from the first-order scalar perturbations.

Next we have derived the second-order temperature anisotropies of CMB
in terms of these perturbations. They will be
useful to analyze the nonlinear influence of local inhomogeneities (at
later stages) upon observed CMB anisotropies. Since nonlinearity
brings the coupling of two linearly independent inhomogeneities with
different wavelengths, it may appear as a
small directional asymmetry and some deviations from the results which
are expected in the standard linear theory.  

\acknowledgments{The author thanks the referee for providing a helpful
comment on vector perturbations.}


\appendix
\section{Second-order perturbed Ricci tensors and Einstein equations}

We show here the expressions for perturbations of Ricci tensors to
derive the perturbed Einstein equations.
For the metric perturbations $\delta g_{\mu\nu} = h_{\mu\nu} +
\ell_{\mu\nu}$, the contravariant metric perturbations $\delta
g^{\mu\nu}$ are derived from the condition $(g_{\mu\nu} +\delta
g_{\mu\nu}) (g^{\nu\gamma} + \delta g^{\nu\gamma}) =
\delta_\mu^\gamma$ as
\begin{equation}
  \label{eq:a1}
\delta g^{\mu\nu} = - h^{\mu\nu} + (h^\mu_\gamma h^{\gamma\nu} -
\ell^{\mu\nu}).
\end{equation}
The perturbed Christoffel symbols are
\begin{eqnarray}
  \label{eq:a2}
\mathop{\delta}_1 \Gamma^\mu_{\nu\gamma} &=& {1 \over 2}
(h^\mu_{\nu;\gamma} + h^\mu_{\gamma;\nu} - {h_{\nu\gamma}}^{;\mu}), \cr
\mathop{\delta}_2 \Gamma^\mu_{\nu\gamma} &=& {1 \over 2}
(\ell^\mu_{\nu;\gamma} + \ell^\mu_{\gamma;\nu} - {\ell_{\nu\gamma}}^{;\mu}) 
- {1 \over 2} h^\mu_\lambda (h^\lambda_{\nu;\gamma} +
h^\lambda_{\gamma;\nu} - {h_{\nu\gamma}}^{;\lambda}), 
\end{eqnarray}
where a semicolon denotes the four-dimensional covariant derivative in
the unperturbed universe. From Eqs. (\ref{eq:a1}) and (\ref{eq:a2}), we
can derive the perturbed curvature tensors and Ricci tensors:
\begin{eqnarray}
  \label{eq:a3}
\mathop{\delta}_1 R^\mu_{\nu\lambda\gamma} &=& {1 \over 2} 
(h^\mu_{\nu;\gamma\lambda} +h^\mu_{\gamma;\nu\lambda} -
{h_{\nu\gamma}}^{;\mu}_{;\lambda} - h^\mu_{\lambda;\nu\gamma} -
h^\mu_{\nu;\lambda\gamma} + {h_{\nu\lambda}}^{;\mu}_{;\gamma}), \cr
\mathop{\delta}_1 R_{\nu\gamma} &\equiv& \mathop{\delta}_1
R^\mu_{\nu\mu\gamma} =  {1 \over 2} 
(h^\mu_{\nu;\gamma\mu} +h^\mu_{\gamma;\nu\mu} -
{h_{\nu\gamma}}^{;\mu}_{;\mu} - h_{;\nu\gamma}), \cr
\mathop{\delta}_2 R^\mu_{\nu\lambda\gamma} &=& {1 \over 2} 
(\ell^\mu_{\nu;\gamma\lambda} +\ell^\mu_{\gamma;\nu\lambda} -
{\ell_{\nu\gamma}}^{;\mu}_{;\lambda} - \ell^\mu_{\lambda;\nu\gamma} -
\ell^\mu_{\nu;\lambda\gamma} + {\ell_{\nu\lambda}}^{;\mu}_{;\gamma}) \cr
&-& {1 \over 2}h^\mu_\alpha (h^\alpha_{\nu;\gamma\lambda} +
h^\alpha_{\gamma;\nu\lambda} - 
{h_{\nu\gamma}}^{;\alpha}_{;\lambda} - h^\alpha_{\lambda;\nu\gamma} -
h^\alpha_{\nu;\lambda\gamma} + {h_{\nu\lambda}}^{;\alpha}_{;\gamma}) \cr
&+& {1 \over 4}(h^\mu_{\alpha;\gamma} - h^\mu_{\gamma;\alpha} +
{h_{\gamma\alpha}}^{;\mu}) (h^\alpha_{\nu;\lambda} +
h^\alpha_{\lambda;\nu} - {h_{\nu\lambda}}^{;\alpha}) \cr  
&-& {1 \over 4}(h^\mu_{\alpha;\lambda} - h^\mu_{\lambda;\alpha} +
{h_{\alpha\lambda}}^{;\mu}) (h^\alpha_{\nu;\gamma}
+h^\alpha_{\gamma;\nu} - {h_{\nu\gamma}}^{;\alpha}), \cr 
\mathop{\delta}_2 R_{\nu\gamma} &\equiv& \mathop{\delta}_2
R^\mu_{\nu\mu\gamma} =  {1 \over 2} 
(\ell^\mu_{\nu;\gamma\mu} +\ell^\mu_{\gamma;\nu\mu} -
{\ell_{\nu\gamma}}^{;\mu}_{;\mu} - \ell_{;\nu\gamma}) \cr
&-& {1 \over 2}h^\mu_\alpha (h^\alpha_{\nu;\gamma\mu} +
h^\alpha_{\gamma;\nu\mu} - 
{h_{\nu\gamma}}^{;\alpha}_{;\mu} - h^\alpha_{\mu;\nu\gamma}) \cr
&+& {1 \over 4}(h^\mu_{\alpha;\gamma} - h^\mu_{\gamma;\alpha} +
{h_{\gamma\alpha}}^{;\mu}) (h^\alpha_{\nu;\mu} +
h^\alpha_{\mu;\nu} - {h_{\nu\mu}}^{;\alpha}) \cr  
&-& {1 \over 4}(2 h^\mu_{\alpha;\mu} - h_{;\alpha}) 
(h^\alpha_{\nu;\gamma}
+h^\alpha_{\gamma;\nu} - {h_{\nu\gamma}}^{;\alpha}).
\end{eqnarray}
Using the relation 
\begin{equation}
  \label{eq:a4}
R^\mu_\nu + \mathop{\delta}_1 R^\mu_\nu + \mathop{\delta}_2 R^\mu_\nu
= (g^{\mu\gamma} + \delta g^{\mu\gamma}) (R_{\nu\gamma}
+\mathop{\delta}_1 R_{\nu\gamma} + \mathop{\delta}_2 R_{\nu\gamma}),
\end{equation}
we obtain the mixed components of the perturbed Ricci tensors:
\begin{eqnarray}
  \label{eq:a5}
\mathop{\delta}_1 R^\mu_\nu &=& g^{\mu\gamma} \mathop{\delta}_1
R_{\nu\gamma} - h^{\mu\gamma} R_{\nu\gamma}, \cr
\mathop{\delta}_2 R^\mu_\nu &=& g^{\mu\gamma} \mathop{\delta}_2
R_{\nu\gamma} - h^{\mu\gamma} \mathop{\delta}_1 R_{\nu\gamma} +
(h^\mu_\lambda h^{\lambda\gamma} - 
\ell^{\mu\gamma}) R_{\nu\gamma}. 
\end{eqnarray}
Here let us impose the synchronous gauge conditions (\ref{eq:m7}) and
(\ref{eq:m16}) on $h_{\mu\nu}$ and
$\ell_{\mu\nu}$. Then we obtain the following expressions for
second-order perturbed Ricci tensors $\mathop{\delta}_2 R^\mu_\nu$.
\begin{eqnarray}
  \label{eq:a6}
2a^2 \mathop{\delta}_2 R^j_i &=& \Phi^j_i +(\ell^j_i)'' + {2a' \over
a}(\ell^j_i)' + {a' \over a}\ell'\delta^j_i - [h^j_k (h^k_i)'' \cr
&+& (h^k_i)'(h^j_k)' - {1 \over 2} h' (h^j_i)' + {a' \over
a}\delta^j_i h^k_l (h^l_k)' + {2a' \over a} h^j_k (h^k_i)'], \cr 
2a^2 \mathop{\delta}_2 R^0_i &=& (\ell_{|i})' - (\ell^k_{i|k})' +
h^k_l [(h^l_{i|k})' - (h^l_{k|i})'] \cr
&-& {1 \over 2} (h^k_l)' h^l_{k|i} + {1 \over 2} (h^k_i)'(2 h^l_{k|l}
- h_{|k}), \cr
2a^2 \mathop{\delta}_2 R^j_0 &=& (\ell^{|j})' -(\ell^{j|k}_k)' +
h^{kl}(h^j_{k|l})' - h^k_l (h^{l|j}_k)' - {1 \over 2}(h^k_l)'
h_k^{l|j} \cr
&+& {1 \over 2}(h^j_k)' (2h_l^{k|l} - h^{|k}) + h^j_k [(h^{|k})' -
(h_l^{k|l})'], \cr
2a^2 \mathop{\delta}_2 R^0_0 &=& \ell'' + {a' \over a}\ell' - h^k_l
[(h^l_k)'' + {a' \over a}(h^l_k)'] - {1 \over 2}(h^k_l)'(h^l_k)', \cr
2a^2 \mathop{\delta}_2 R &\equiv& 2a^2 \mathop{\delta}_2 (R^0_0 +
R^k_k) = \Phi^k_k + 2(\ell'' + 3 {a' \over a}\ell') \cr
&-& [2h^k_l (h^l_k)'' + {3 \over 2}(h^k_l)'(h^l_k)' - {1 \over
2}(h')^2 + {6a' \over a} h^k_l (h^l_k)'] ,
\end{eqnarray}
where $|i$ means the three-dimensional covariant derivative in the
constant-curvature space with
metric $\gamma_{ij}$, $ h = h^k_k, \ \ell = \ell^k_k$ \ and
$\Phi^j_i$ is defined as follows: 
\begin{eqnarray}
 \label{eq:a7}
\Phi^j_i &\equiv& {\ell^{k|j}_i}_{|k} + {\ell^j_{k|i}}^{|k} -
{\ell_i^{j|k}}_{|k} - \ell^{|k}_{|k} - 4K \ell^j_i \cr
&-& h^k_l [{h^{l|j}_i}_{|k} + {h^j_{k|i}}^{|l} - {h^{j|l}_i}_k-
{h^l_{k|i}}^{|j}] \cr
&-& h^j_l [{h^{k|l}_i}_{|k} + {h^l_{k|i}}^k - {h^{l|k}_i}_{|k} -
h^{|l}_{|i} - 4K h^l_i] \cr
&+& {1 \over 2} (h^{k|j}_l - \gamma^{km}h^j_{m|l} + h_l^{j|k})
(h^l_{i|k} +h^l_{k|i} - \gamma_{mk} h_i^{m|l}) \cr
&-& {1 \over 2}(2h^k_{l|k} - k_{|l}) ( h^{l|j}_i + \gamma^{lm}
h^j_{m|i} - h_i^{j|l}),
\end{eqnarray}
where $K\ (=\pm 1, 0)$ is the three-dimensional spatial curvature.
In the Appendix A, we take the spatial curvature and the pressure $p$
into consideration. 

The first-order and second-order perturbed Einstein equations are
expressed as
\begin{eqnarray}
 \label{eq:a8}
\mathop{\delta}_1 R^\mu_\nu &-& {1 \over 2} \delta^\mu_\nu
\mathop{\delta}_1 R = \mathop{\delta}_1 T^\mu_\nu, \cr
\mathop{\delta}_2 R^\mu_\nu &-& {1 \over 2} \delta^\mu_\nu
\mathop{\delta}_2 R = \mathop{\delta}_2 T^\mu_\nu.
\end{eqnarray}
The perturbations of the energy-momentum tensor are
\begin{eqnarray}
 \label{eq:a9}
\mathop{\delta}_1 T^\mu_\nu &=& \delta^\mu_\nu \mathop{\delta}_1 p +
g_{\nu\alpha} [u^\alpha u^\mu (\mathop{\delta}_1 p + \mathop{\delta}_1
\rho) +(\mathop{\delta}_1 u^\alpha u^\mu + u^\alpha \mathop{\delta}_1
u^\mu)(p + \rho)] \cr
&+& h_{\nu\alpha} u^\alpha u^\mu (p + \rho), \cr
\mathop{\delta}_2 T^\mu_\nu &=& \delta^\mu_\nu \mathop{\delta}_2 p +
g_{\nu\alpha} [u^\alpha u^\mu (\mathop{\delta}_2 p + \mathop{\delta}_2
\rho) +(\mathop{\delta}_1 u^\alpha u^\mu + u^\alpha \mathop{\delta}_1
u^\mu)(\mathop{\delta}_1 p + \mathop{\delta}_1 \rho) \cr
&+& (\mathop{\delta}_2 u^\alpha u^\mu + u^\alpha \mathop{\delta}_2 
u^\mu + \mathop{\delta}_1 u^\alpha \mathop{\delta}_1 u^\mu)(p + \rho)] \cr
&+& h_{\nu\alpha} [u^\alpha u^\mu (\mathop{\delta}_1 p +
\mathop{\delta}_1 \rho) + (\mathop{\delta}_1 u^\alpha u^\mu + u^\alpha
\mathop{\delta}_1 u^\mu) (p + \rho)] \cr
&+& \ell_{\nu\alpha} u^\alpha u^\mu (p + \rho). 
\end{eqnarray}

For velocity perturbations, we obtain
\begin{equation}
  \label{eq:a10}
\mathop{\delta}_1 u^0 = 0, \ \ \mathop{\delta}_2 u^0 = {1 \over 2a}
g_{kl} \mathop{\delta}_1 u^k \mathop{\delta}_1 u^l, 
\end{equation}
using the identity
\begin{equation}
  \label{eq:a10a}
(g_{\mu\nu} +h_{\mu\nu} + \ell_{\mu\nu})(u^\mu +\mathop{\delta}_1
u^\mu +\mathop{\delta}_2 u^\mu)(u^\nu +\mathop{\delta}_1 u^\nu
+\mathop{\delta}_2 u^\nu) = -1 
\end{equation}
and the synchronous gauge conditions. The components of the
energy-momentum tensor are, therefore, 
\begin{equation}
 \label{eq:a11}
\mathop{\delta}_1 T^j_i = \delta^j_i \mathop{\delta}_1 p, \ \
\mathop{\delta}_1 T^j_0 = -a(p + \rho)\mathop{\delta}_1 u^j, \ \
\mathop{\delta}_1 T^0_0 = - \mathop{\delta}_1 \rho, 
\end{equation}
and
\begin{eqnarray}
 \label{eq:a12}
\mathop{\delta}_2 T^j_i &=& \delta^j_i \mathop{\delta}_2 p + (p+\rho)
g_{ik} \mathop{\delta}_1 u^k \mathop{\delta}_1 u^j, \cr
\mathop{\delta}_2 T^0_i &=& {1 \over a} g_{ik}[\mathop{\delta}_1 u^k
(\mathop{\delta}_1 p + \mathop{\delta}_1 \rho)] + \mathop{\delta}_2 u^k
(p+\rho) + {1 \over a} h_{ik} \mathop{\delta}_1 u^k (p+ \rho), \cr
\mathop{\delta}_2 T^j_0 &=& -a \mathop{\delta}_1 u^j
(\mathop{\delta}_1 p + \mathop{\delta}_1 \rho) - a \mathop{\delta}_2
u^j (p + \rho), \cr
\mathop{\delta}_2 T^0_0 &=& - \mathop{\delta}_2 \rho - (p+ \rho)
g_{kl} \mathop{\delta}_1 u^k \mathop{\delta}_1 u^l. 
\end{eqnarray}
If we assume the equation of state $p = p(\rho)$, then we have
\begin{equation}
  \label{eq:a13}
\mathop{\delta}_1 p = (dp/d \rho) \mathop{\delta}_1 \rho \ \ {\rm and
} \ \ \mathop{\delta}_2 p = (dp/d \rho) \mathop{\delta}_2 \rho + {1
\over 2} (d^2 p/d\rho^2) (\mathop{\delta}_1 \rho)^2,
\end{equation}
so that
\begin{eqnarray}
  \label{eq:a14}
\mathop{\delta}_1 T^j_i &=& - \delta^j_i (dp/d\rho) \mathop{\delta}_1
T^0_0, \cr
\mathop{\delta}_2 T^j_i &=& - \delta^j_i (dp/d\rho) \mathop{\delta}_2
T^0_0 + {1 \over 2} \delta^j_i (d^2 p/d\rho^2) (\mathop{\delta}_1
\rho)^2 \cr
&+& (p + \rho) [g_{ik} \mathop{\delta}_1 u^j - \delta^j_i (dp/d\rho)
g_{kl} \mathop{\delta}_1 u^l] \mathop{\delta}_1 u^k.
\end{eqnarray}
When the first-order quantities $h^j_i, \mathop{\delta}_1 u^i,$ and
$\mathop{\delta}_1 \rho$ are known, the equations for determining the
second-order metric perturbations are derived from Eqs.(\ref{eq:a8})
and (\ref{eq:a14}) as follows:
\begin{eqnarray}
  \label{eq:a15}
(\ell^j_i)'' &+& {2a' \over a} (\ell^j_i)' + \Phi^j_i = h^j_k
(h^k_i)'' + (h_i^k)'(h_k^j)' - {1 \over 2} h' (h^j_i)'  \cr
&+& {2a' \over a} h^j_k (h^k_i)' + 2a^2 (p+ \rho) g_{ik}
\mathop{\delta}_1 u^k \mathop{\delta}_1 u^j, \ \ (i \ne j), 
\end{eqnarray}
\begin{eqnarray}
  \label{eq:a16}
(\ell^i_i)'' &+& {2a' \over a} (\ell^i_i)'  - \ell'' -  {2a' \over a} \ell'
(1 + dp/d\rho) + \Phi^i_i - {1 \over 2} \Phi^k_k (1 + dp/d\rho) \cr
&=&  h^i_k (h^k_i)'' + (h_i^k)'(h_k^i)' - {1 \over 2} h' (h^i_i)'
+{2a' \over a} h^i_k (h^k_i)' \cr 
&-& h^l_k (h^k_l)'' - {3 \over 4} (h_l^k)'(h_k^l)' + {1 \over 4} (h')^2
-{2a' \over a} h^l_k (h^k_l)' \cr
&+& 2a^2 (p+ \rho) \Bigl[g_{ik} \mathop{\delta}_1 u^i - (dp/d\rho) g_{kl}
\mathop{\delta}_1 u^l \Bigr] \mathop{\delta}_1 u^k \cr
&-& (dp/d\rho) [{1 \over 4} (h_l^k)'(h_k^l)' +{2a' \over a} h^i_k
(h^k_i)' - (h')^2]  \cr 
&+& a^2 (d^2 p/d\rho^2) (\mathop{\delta}_1 \rho)^2,
\end{eqnarray}
where in the latter equation the index $i$ is non-dummy but others are
dummy, \ $h = h^k_k, 
\ \ell = \ell^k_k$, \ and $\Phi^j_i$ is defined in Eq.(\ref{eq:a7}).
Eqs.(\ref{eq:a15}) and (\ref{eq:a16}) consist of six equations,
which corresponds to six independent components of $\ell^j_i$.

The second-order density and velocity perturbations can be expressed
in terms of metric perturbations as
\begin{eqnarray}
  \label{eq:a17}
\mathop{\delta}_2 \rho/\rho &=& {1 \over \rho}\Bigl[-\mathop{\delta}_2
R^0_0 + {1 
\over 2} \mathop{\delta}_2 R - (p+ \rho) g_{kl} \mathop{\delta}_1 u^k
\mathop{\delta}_1 u^l \Bigr] \cr
&=& {1 \over 2\rho a^2}\Bigl[{2a' \over a}\ell' - {1 \over
4}(h_l^k)'(h_k^l)' - 
{2a' \over a}h_l^k (h_k^l)' + {1 \over 4}(h')^2 \cr
&+& {1 \over 2} \Phi^k_k - 2a^2 (p+ \rho)g_{kl} \mathop{\delta}_1 u^k
\mathop{\delta}_1 u^l \Bigr], 
\end{eqnarray}
\begin{eqnarray}
  \label{eq:a18}
a \mathop{\delta}_2 u^i &=& - {1 \over p+ \rho}
\Bigl[\mathop{\delta}_2 R^i_0 + a \mathop{\delta}_1 u^i
(\mathop{\delta}_1 p +\mathop{\delta}_1 \rho) \Bigr] \cr
&=& {1 \over 2(p + \rho)a^2}\Big\{(\ell^{|i})' - (\ell^{i|k}_k)' + h^k_l
(h^{i|l}_k)' \cr
&-& h^k_l (h^{l|i}_k)' -{1 \over 2} (h^k_l)' h^{l|i}_k +{1 \over 2}
(h^i_l)'(2h_k^{l|k} - h^{|l}) \cr
&+& h^i_k [(h^{|k})' - (h^{k|l}_l)'] + 2a^3 \mathop{\delta}_1 u^i
(\mathop{\delta}_1 p +\mathop{\delta}_1 \rho) \Big\}.
\end{eqnarray}
%



\end{document}